\documentstyle[11pt,epsfig]{article}

\def\be{\begin{eqnarray}}
\def\ee{\end{eqnarray}}

\def\beitem{\begin{itemize}}
\def\eitem{\end{itemize}}
\def\ben{\begin{enumerate}}
\def\een{\end{enumerate}}

\def\roughly#1{\mathrel{\raise.3ex\hbox{$#1$\kern-.75em%
    \lower1ex\hbox{$\sim$}}}}

\newcommand{\beq}{\begin{eqnarray}}
\newcommand{\eeq}{\end{eqnarray}}

\def\bi{\begin{itemize}}
\def\ei{\end{itemize}}
\def\ba{\begin{array}}
\def\ea{\end{array}}

\long\def\beginomit#1\endomit{}

\begin{document}



\hfill{SUNY-NTG-97-10}


\hfill{\today}

\vskip 0.4in
\begin{center}
{\Large\bf Symmetry Energy of Nuclear Matter\\
 and Properties of
Neutron Stars in a Relativistic Approach} 
\vskip 0.8in
{\large  C.-H. Lee, T.T.S. Kuo, G.Q. Li and G.E. Brown}\\
{\large  \it Department of Physics, State University of New York,} \\
{\large \it Stony Brook, N.Y. 11794, USA. }\\
\vskip 0.4in

{\bf ABSTRACT}\\ \vskip 0.1in
\begin{quotation}

Asymmetric nuclear matter is treated in the formalism of 
Dirac-Brueckner approach with Bonn one-boson-exchange
nucleon-nucleon interaction. We extract the symmetry energy 
coefficient at the saturation to be about 31 MeV, which is in good
agreement with  empirical value of $30\pm 4$ MeV.
The symmetry energy is found to increase almost linearly
with the density, which differs considerably from the results
of non-relativistic approaches. This finding also supports
the linear parameterization of Prakash, Ainsworth and Lattimer.
We find, furthermore, that the higher-order 
dependence of the nuclear equation of state on the
asymmetry parameter is unimportant up to densities relevant
for neutron stars. The resulting equation
of state of neutron-rich matter is used to calculate the
maximum mass of neutron star, and we find it to be about 2.1$M_\odot$.
Possible mechanisms for the softening of the equation of
state are also discussed. 
\end{quotation}
\end{center}


\section{Introduction}

Although the fact that the equation of state of nuclear matter
contains a symmetry energy term has been known since the early
days of nuclear physics, the experimental and theoretical
study of the symmetry energy and its density dependence
is becoming an increasingly interesting topic, mainly because
of the recent development of radioactive ion beam facilities that 
allow one to study the structure and reactions of neutron-rich 
nuclei \cite{tani95,han95,tani96}, in which the symmetry 
energy plays an important role. The recognization that the 
symmetry energy, especially its density dependence, has a 
profound effect on the properties of neutron stars 
\cite{MPA,pra88,lat91,eng94,elg96} also makes the experimental 
and theoretical determination of this quantity very relevant 
and useful.

Empirically, the symmetry energy coefficient $S_2(\rho_0)$ in nuclear matter
at the saturation density $\rho_0$ can be extracted from the
systematic study of the masses of atomic nuclei, based on,
e.g., the liquid droplet model \cite{myers69,myers76} or
the macroscopic-microscopic model \cite{moll81,moll88}.
This, however, determines the symmetry energy only for small 
asymmetry parameter $\alpha$ ($\alpha = (N-Z)/A$) and for densities
around $\rho_0$. From the experience with 
symmetric nuclear matter we know that the determination of the 
compression modulus K at $\rho_0$ does not uniquely constrain the  
equation of state at high densities. Similarly, the determination of
$S_2(\rho_0)$ does not guarantee an unambiguous determination of
symmetry energy at high densities which is needed for
the study of neutron star properties.  

The situation changes with the recent advances in the 
development of various radioactive ion beam facilities 
around the world that will produce nuclei with large neutron excess
near and beyond the drip-line. The study of the structure
of these neutron-rich nuclei allows us to determine
the symmetry energy for large asymmetry parameter and 
extract possible higher-order dependence on $\alpha$.
Furthermore, the collisions of neutron-rich nuclei
at relativistic energies, during which nuclear matter with
densities up to (2-3)$\rho_0$ is created, make it possible to
study experimentally the density dependence of the symmetry energy
\cite{pak,bali}.

On the theoretical side, the symmetry energy has been studied
over many years based on various models and approaches,
which can roughly be divided into phenomenological and
microscopic, each of which can be subdivided into relativistic
and non-relativistic. Hartree-Fock \cite{hf} and Thomas-Fermi \cite{tf}
calculations with Skyrme-type effective nucleon-nucleon
interactions have been carried out to study various
nuclear properties, including symmetry energy. 
The symmetry energy coefficient $S_0$ from these
calculations ranges from about 27 to 38 MeV, and
is in agreement with the empirical value of $30\pm 4$ MeV \cite{edata}.  
Since the the parameters in the Skyrme forces are adjusted to
fit nuclear matter and finite nuclei properties,
this degree of agreement is fully expected.

Another phenomenological approach that has been used extensively in the 
study of nuclear properties is quantum hadrodynamics (QHD)
which is based on the relativistic field theory \cite{qhd,serot}.
The symmetry energy in this approach has two contributions
\cite{serot79,mat81}; one comes from the `kinetic' energy 
difference between symmetric and asymmetric matter, and
the other arises from the exchange of the rho meson which
couples to neutron and proton with opposite sign.
The symmetry energy in this approach ranges from about 35 to 40 MeV
\cite{mat81,rufa88,funs95}, somewhat larger than the emprical
value of $30\pm 4$ MeV.

On a more microscopic level, it has been the ultimate goal
of traditional nuclear physics to describe in a consistent
way the properties of nuclear matter, finite nuclei, and
nuclear reactions from realistic nucleon-nucleon interactions that are
fitted to the nucleon-nucleon scattering and deuteron data. 
There are basically two approaches to this; namely, variational-type
and Brueckner-type. In variational calculations, it is well-known
that the realistic two-nucleon potential alone, such as Argonne
$v_{14}$ (AV14) or Urbana $v_{14}$ (UV14), does not provide a 
satisfactory description of nuclear matter properties \cite{pan81,wir88}. 
A phenomenological three-nucleon interaction has to be
introduced, whose parameters are adjusted so that the variational
calculations with the two- and three-body interactions give 
correct nuclear matter saturation density, binding energy and
compression modulus. The symmetry energy coefficient obtained in the
variational calculations is about 30 MeV \cite{pan81,wir88}.

The Brueckner-type approach has both non-relativistic and
relativistic versions, known as Brueckner-Hartree-Fock (BHF) 
\cite{ bom91,song92} and Dirac-Brueckner-Hartree-Fock (DBHF)
\cite{shakin83,mal87,mach89,brock90,li92}, respectively. 
As in the variational calculations, the BHF calculation with 
realistic two-nucleon interactions such as Bonn and Paris 
potentials does not provide a good description of nuclear matter 
properties \cite{bom91}. On the other hand, the DBHF approach, 
owning to the additional density dependence introduced through 
the medium modified Dirac spinors, does provide a very good 
description of nuclear matter properties based entirely on 
realistic nucleon-nucleon interactions that are constrained 
by two-nucleon data (nucleon-nucleon scattering and deuteron properties). 
  
In this work, we study systematically the properties of 
asymmetric nuclear matter in the formalism of
the DBHF approach using the Bonn one-boson-exchange (OBE) potential.
We will calculate in particular the density dependence of
the symmetry energy that is very important for neutron star
properties and heavy-ion collisions. Our results will be compared
to those of the variational and BHF calculations,
and to various phenomenological parameterizations. 
We will also apply the nuclear equation of state obtained 
in this study to calculate neutron star properties.
In Section 2, we review briefly the DBHF approach and
its extension to asymmetric nuclear matter. The
results for symmetric energy will be presented in
Section 3, while those for neutron stars in Section 4.
A short summary is given in Section 5.

\section{Dirac Brueckner approach for asymmetric nuclear matter}

The essential point of the DBHF approach is the use of the Dirac
equation for the description of the single-particle motion in
the nuclear medium. The Dirac spinor, which enters the
evaluation of in-medium nucleon-nucleon potential, 
becomes density dependent. This additional density
dependence is instructive in reproducing correctly the
nuclear matter saturation density and binding energy \cite{mach89}.

The basic quantity in the DBHF calculation is the $\tilde G$
matrix which satisfies the in-medium Thompson equation,
\be
\tilde G (q^\prime,q | P,\tilde z) &=&
\tilde V (q^\prime,q )
+\int \frac{d^3k}{(2\pi)^3} \tilde V (q^\prime,k )
  \left(\frac{\tilde m (k)}{\tilde E(k)}\right)^2
  \frac{\bar Q(k,P)}{2\tilde E(q) -2 \tilde E (k) }
\tilde G (k,q | P,\tilde z)
\ee
where $\tilde E = \sqrt {\tilde m^2 + ({\bf P}/2+{\bf k})^2}$, and
$\tilde m=m+U_S$, with $m$ being nucleon mass in free space. 
For asymmetric nuclear matter, the angle-averaged Pauli-blocking
operator has to be modified and is given by
\be
\bar Q(k,K) &=& \left\{
\begin{array}{lll}
1                   & {\rm if} & \beta_n > 1 \\
(1+\beta_n)/2       & {\rm if} & -1 < \beta_n < 1 \; {\rm and} \; \beta_p >1 \\
(\beta_n+\beta_p)/2 & {\rm if} & \beta_p < 1 \;\; {\rm and}\;\;
  0< (\beta_n+\beta_p)/2 \\
 0                  & {\rm if} & (\beta_n+\beta_p)/2 <0 \; {\rm or} \; \beta_n < -1\\
\end{array}\right.  ,
\ee
where
\be
\beta_{n,p} = \frac{K^2/4+k^2-k_{F_{n,p}}^2}{Kk}
\ee
where $k_{F_{n}}$ and $k_{F_{p}}$ are neutron and proton Fermi
momenta, respectively, with $k_{F_n} \ge k_{F_p}$.

{}From the $\tilde G$ matrix we can calculate the single-particle
potential
\be
\Sigma_{DBHF}(k) &=& Re \int_0^{k_F} d^3q
  \left(\frac{\tilde m(q)}{\tilde E(q)}\right)
  \left(\frac{\tilde m(k)}{\tilde E(k)}\right)
 \left\langle k q \left| \tilde G (\tilde z) \right| kq-qk\right\rangle ,
\ee
which is usually parameterized in terms of the scalar and vector
potentials
\be
\Sigma (k) &=& \frac{\tilde m(k)}{\tilde E (k)} U_S (k) +U_V (k).
\label{sigmaf}
\ee
In principle, the scalar and vector potentials are both density and
momentum dependent. The momentum dependence for momenta
below the Fermi momentum is rather weak and smooth 
\cite{mal87,mach89,lee97}. In a recent work \cite{lee97},
we found that the saturation properties and nuclear equation of
state of symmetric matter with explicit momentum dependence 
are almost the same as those obtained in Ref. \cite{mach89}
under the assumption of momentum independent mean fields.
In this work we thus follow Ref. \cite{mach89} and neglect the
momentum dependence of the scalar and vector potentials.

We will use the Bonn A potential from Ref. \cite{mach89}. The
parameters in this potential are fitted to neutron-proton ($np$)
scattering and deuteron properties. It is well known that at low
energies, there are both charge independence breaking
and charge symmetry breaking \cite{mach89}, namely,
{\it np, pp,} and {\it nn} potentials are not completely
identical.  The effects of those differences on the bulk nuclear
matter properties are, however, found to be extremely small 
and can well be neglected \cite{li93}.

In the case of asymmetric nuclear matter, the potential energy 
of a single particle is
\be
E_{pot} &=&\frac{1}{\int_0^{k_{F_n}} d^3k+\int_0^{k_{F_p}}d^3k} 
 \left(\int_0^{k_{F_n}} d^3k \frac 12 \Sigma_n (k)
 +\int_0^{k_{F_p}} d^3k \frac 12 \Sigma_p (k)
\right),
\ee
while the kinetic energy is given by
\be
E_{kin} &=&\frac{1}{\int_0^{k_{F_n}} d^3k+\int_0^{k_{F_p}}d^3k} 
\left(\int_0^{k_{F_{n}}} d^3k
\frac{m m^\star(k) +k^2}{E^\star(k)} 
 +\int_0^{k_{F_{p}}} d^3k
\frac{m m^\star(k) +k^2}{E^\star(k)}
\right), 
\ee
The energy per nucleon, or nuclear equation of state, is then given by
\be
E= E_{pot} +E_{kin} -m.
\ee

\begin{figure}
\centerline{
\epsfig{file=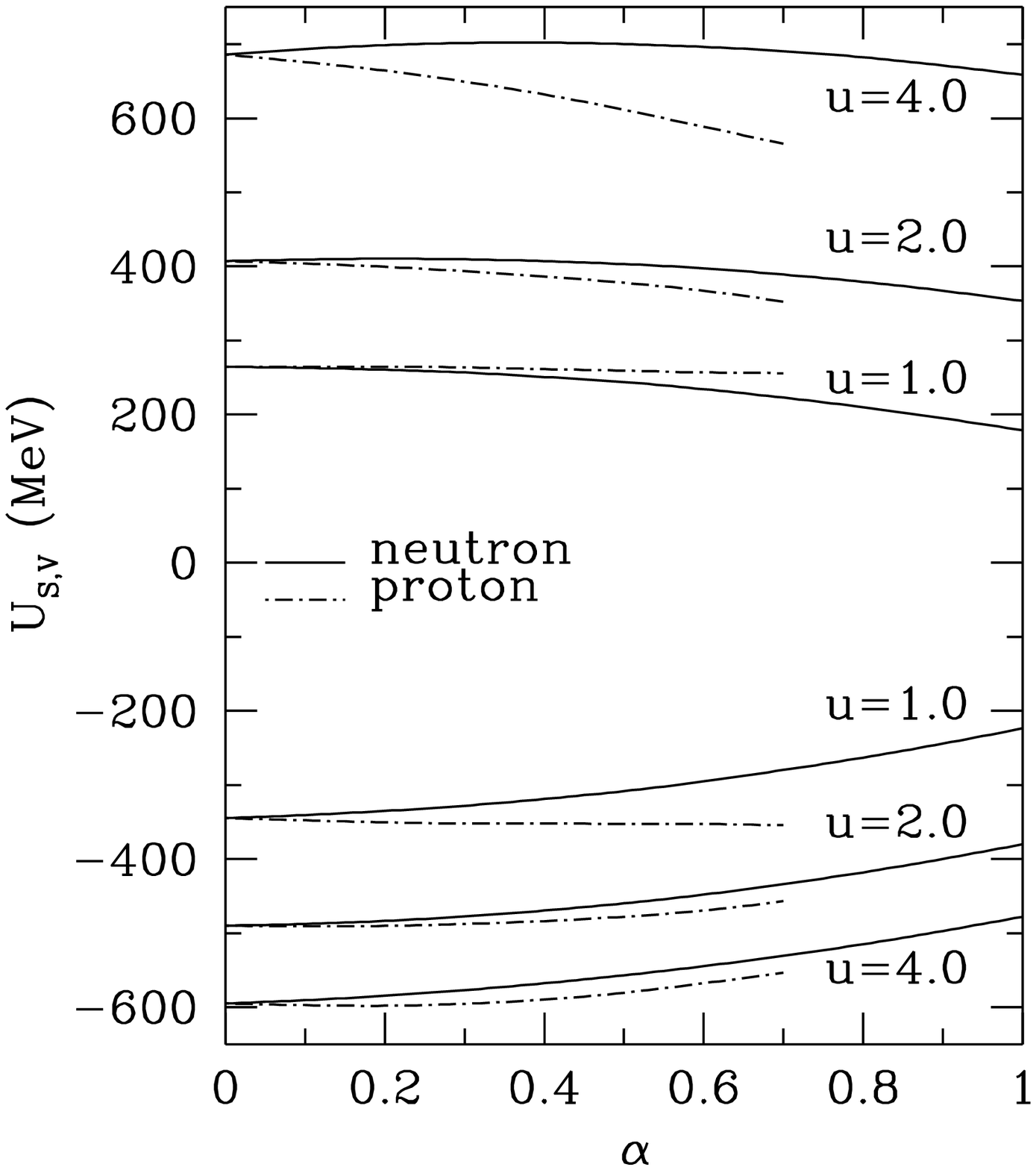,height=10cm}  
          \epsfig{file=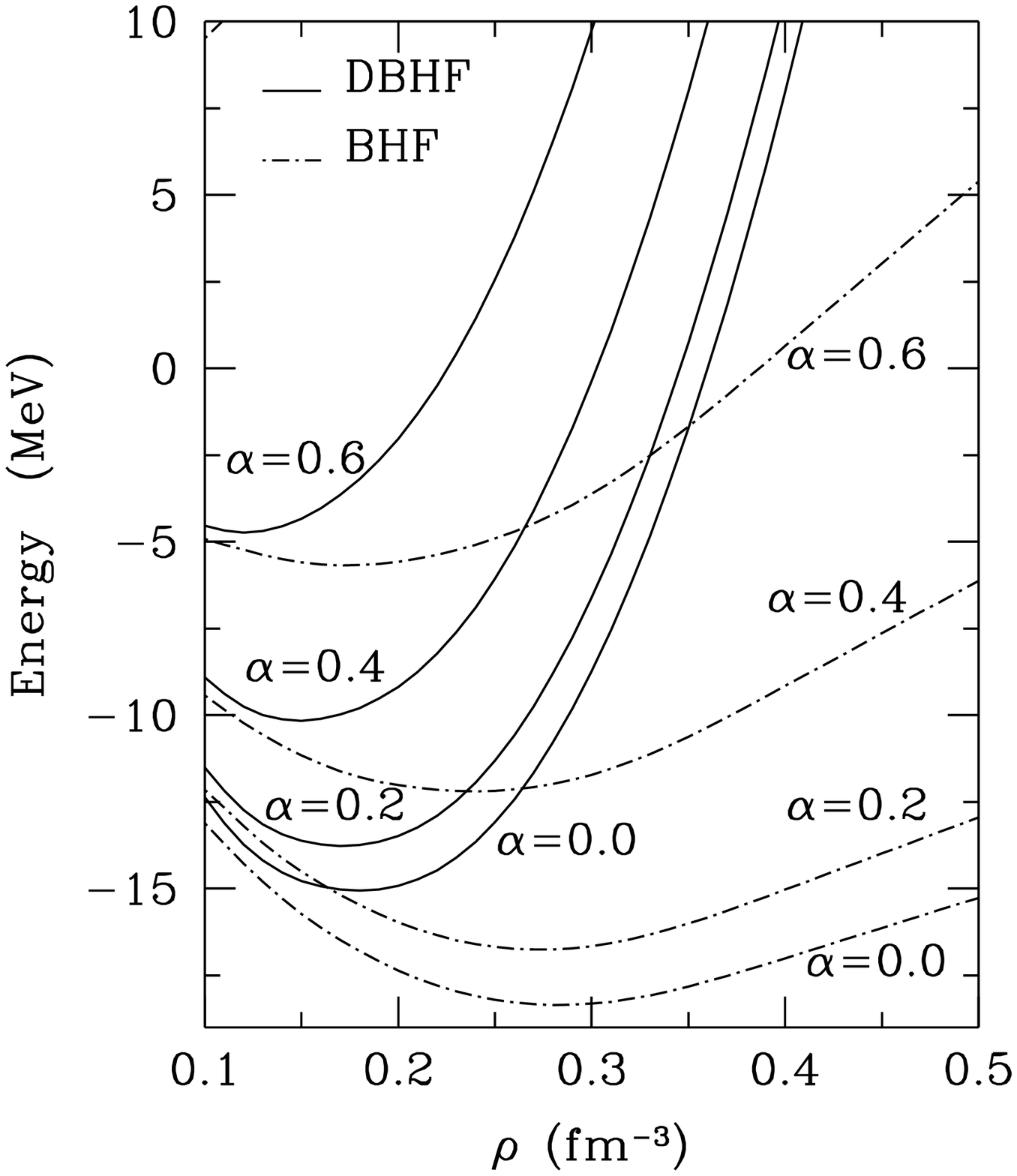,height=10cm} }
\caption{Left panel: neutron and proton scalar and vector potentials
as a function of asymmetry parameter, for three densities.
Right panel: equation of state of nuclear matter for a number
of asymmetry parameters.}
\label{fig1}
\end{figure}

The asymmetric nuclear matter is 
characterized by the asymmetry parameter 
\be
\alpha = \frac{\rho_n-\rho_p}{\rho_n+\rho_p}.
\ee
In the left panel of Fig. \ref{fig1} the neutron and proton scalar 
and vector potentials
are shown as a function of the asymmetry parameter, for three
different densities. It is seen that the variation of the proton and
neutron potentials is not symmetric with respect to their
common value at $\alpha =0$. For a nucleon at rest in nuclear matter,
we can define its single-particle potential as $U_S+U_V$. We find that
as the asymmetry parameter increases, the neutron potential becomes
less attractive and that of proton becomes slightly more attractive.
These differences in proton and neutron potentials may be observed in
heavy-ion collisions with radioactive ion beams \cite{pak,bali}.

In the right panel of Fig. \ref{fig1}, we show the nuclear 
equation of state for a number of asymmetry parameters. 
We compare our results with those of Ref. \cite{bom91} 
obtained in the BHF approach. As is well-known, 
the BHF approach saturates nuclear matter at
a much too high density. In the DBHF calculation,
the nuclear matter saturation
properties are better reproduced. The binding energy and the
saturation density become progressively smaller as $\alpha$ 
increases.

\section{Symmetry Energy of nuclear matter}

\begin{figure}
\centerline{\epsfig{file=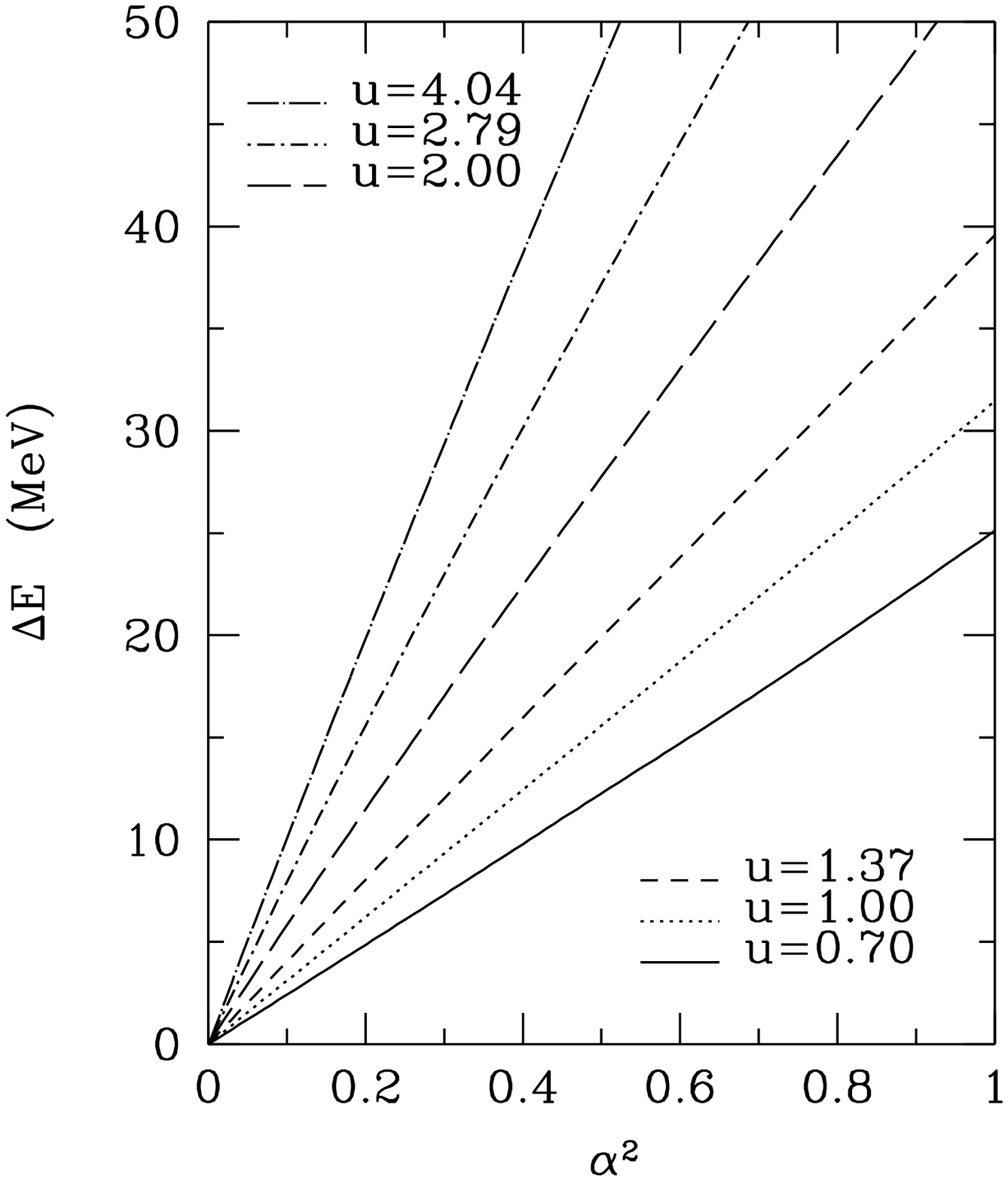,height=10cm}
            \epsfig{file=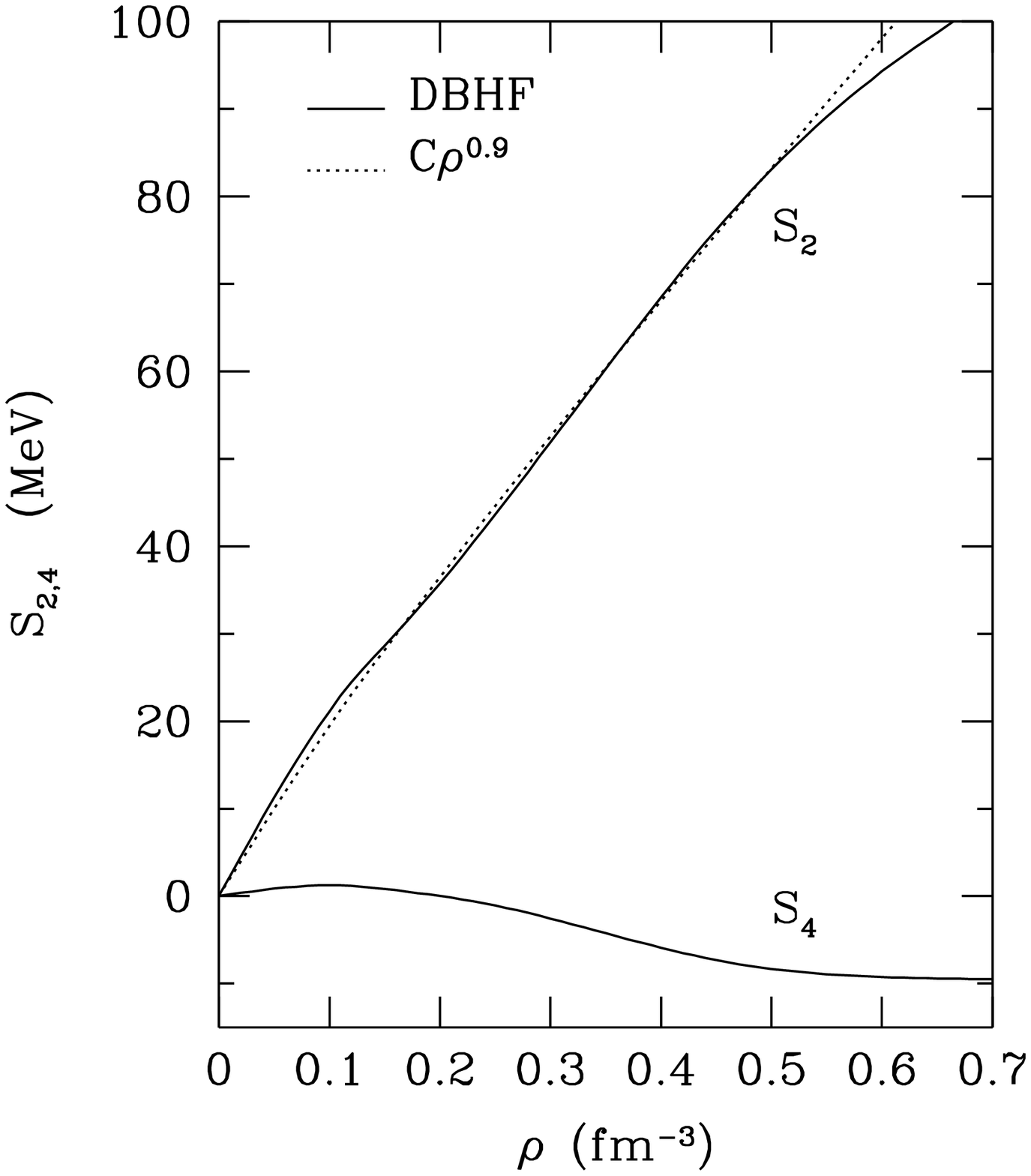,height=10cm}}
\caption{Left panel: energy difference $\Delta E$ as a function of
$\alpha^2$ for several densities with $u=\rho/\rho_0$ and
$\rho_0 = 0.166$ fm$^{-3}$.  Right panel: density dependence
of symmetry parameters.}
\label{fig2}
\end{figure}

In this section, we discuss our results for symmetry energy and 
its density dependence. We introduce $\Delta E$ as the energy 
difference between symmetric and asymmetric nuclear matter,
\be
\Delta E = E(\rho ,\alpha )-E(\rho ,0),
\label{eqde}\ee
with $E(\rho ,\alpha )$ defined in Eq. (7).
The results are shown in the left panel of Fig. \ref{fig2}
as a function of $\alpha ^2$ for six different densities.
It is seen that at all densities considered here, $\Delta E$
increases almost linearly with $\alpha ^2$, indicating that
the $\alpha^4$ and higher-order terms are not important.
To a good extent we can express the equation of
state of asymmetric nuclear matter as
\be
E(\rho ,\alpha )=E(\rho ,0)+S_2(\rho )\alpha ^2 +S_4(\rho )\alpha ^4.
\ee
The usual symmetry energy $S_2$ is thus defined as
\be
S_2 (\rho) &=& \left. 
\frac 12 \frac{\partial^2 {E(\rho ,\alpha )}_{bin}}{\partial 
\alpha^2} \right|_{\alpha=0},
\ee
similarly, 
\be
S_4 (\rho) &=& \left. 
\frac 1{24} \frac{\partial^4 {E(\rho ,\alpha )}_{bin}}{\partial 
\alpha^4} \right|_{\alpha=0},
\ee

The density dependences of $S_2$ and $S_4$ obtained 
in our calculation are shown in the right panel of Fig. 
\ref{fig2} by the solid curve. $S_2$ increases almost linearly
with density. Actually a parameterization in terms of
$(\rho /\rho_0)^{0.9}$ fits the theoretical curve reasonably
well, as shown in the figure by dotted curve. At nuclear 
matter saturation density, our calculation gives a symmetry 
energy coefficient $S_2(\rho _0)$ of about 31 MeV, which 
is in good agreement with the empirical value of 
about $30\pm 4$ MeV \cite{edata}. The BHF and the variational 
calculations also reproduce the empirical symmetry energy 
coefficient \cite{wir88,bom91}. The coefficient of $\alpha^4$
term is very small in the density region considered here.
This means that the approximation of neglecting this
term as adopted in Ref. \cite{pan81,wir88} is quite reasonable.

\begin{figure}
\centerline{\epsfig{file=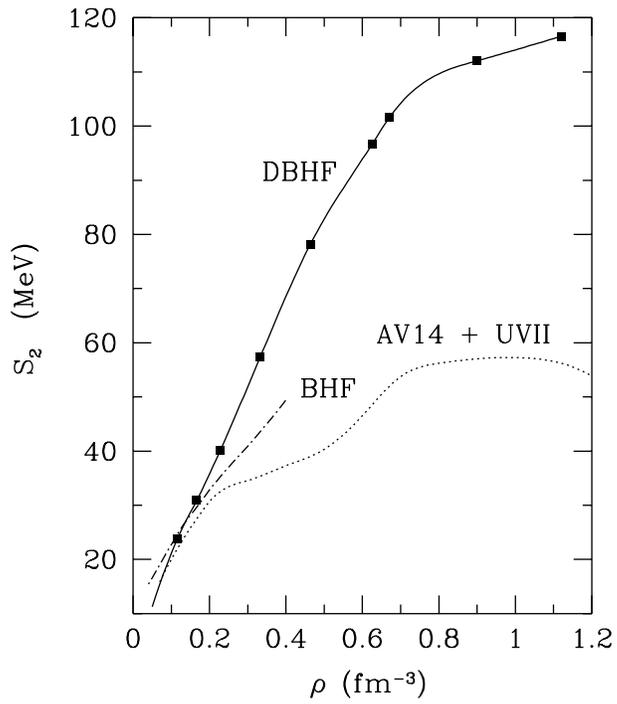,height=10cm}
            \epsfig{file=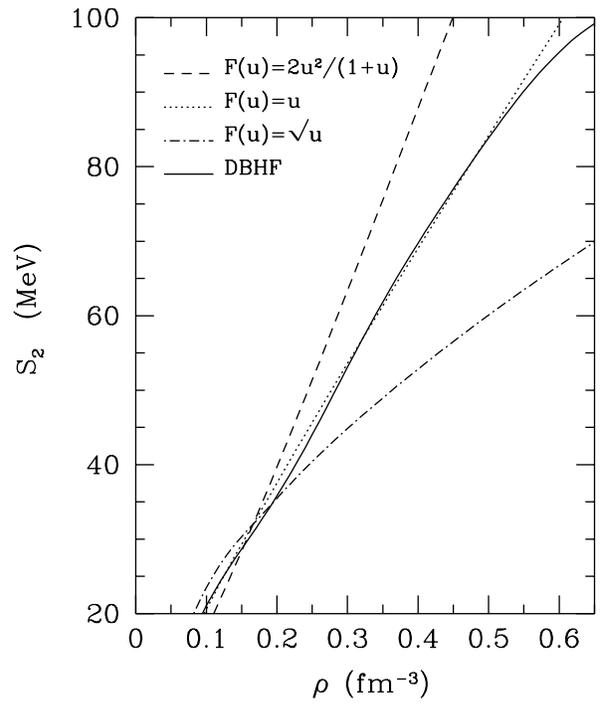,height=10cm}}
\caption{Left panel: comparison of our results with 
those of Refs. \protect\cite{wir88} and \protect\cite{bom91}. 
Right panel: comparisons of our results
with phenomenological parameterizations of 
Prakash, Ainsworth and Lattimer \protect\cite{pra88}.
}
\label{fig3}
\end{figure}

In the left panel of Fig. \ref{fig3} we compare our results 
for the density dependence of the symmetry energy with those of
Ref. \cite{bom91} based on the BHF calculations, and of Ref.
\cite{wir88} based on the variational calculation. 
There are significant differences between the results of 
these three calculations. In relativistic approaches 
\cite{lop88,huber93}, the symmetry energy increases almost 
linearly with density, and is considerably larger than
those in non-relativistic and variational calculations 
\cite{wir88,bom91}.  The difference between the DBHF and BHF
is mainly due to the relativitic effects. In the simple
mean-field approximation to the Walecka-type model, the
symmetry energy has a contribution from the 'kinetic
energy' difference, which is inversely proportional to
$E_F^* = \sqrt {k_F^2+m^{*2}}$. This contribution is thus
larger in relativistic approaches because of the dropping
nucleon mass. This also accounts for part of the 
difference between our results and that of Ref. \cite{wir88}
which is non-relativistic in nature.
The remaining difference can be explained by the differences
in the nucleon-nucleon potentials used in the two calculations.
In the variational calculations \cite{wir88}, the major contribution 
to the 'potential' part of the sysmetry energy comes chiefly from
the second-order tensor interaction, which is progressively 
blocked  with increasing density.
With the strong $\rho$-coupling of the Bonn potential,
the second-order tensor force is relatively weak, compared with that
of Ref. \cite{wir88}, so that this is not a large effect in our calculation,
where the main contribution to the symmetry energy comes from 
$\rho$-meson exchange. The differences in the
symmetry energy in these three calculations will have a profound 
impact on the properties of neutron stars. We hope that future experiments
with radioactive ion beams will help to shed light on this
problem.

Phenomenologically, Prakash, Ainsworth and Lattimer \cite{pra88} 
proposed the following
parameterization for the density dependence of the symmetry energy,
\be
S(u) =(2^{2/3}-1)\frac 35 E_F^0 (u^{2/3}-F(u))+S_0 F(u),
\ee
with $F(u)=2u^2/(1+u)$, $u$, or $\sqrt u$, where $u=\rho/\rho_0$ and
$E_F^0$ is Fermi energy at saturation density $\rho_0$.
In the right panel of Fig. \ref{fig3}, the density dependences 
of three forms of $F(u)$ are compared with our results, and
it is seen that the $F(u)=u$ case is very close to our results.

\section{Nucleon star properties}

\begin{figure}
\centerline{\epsfig{file=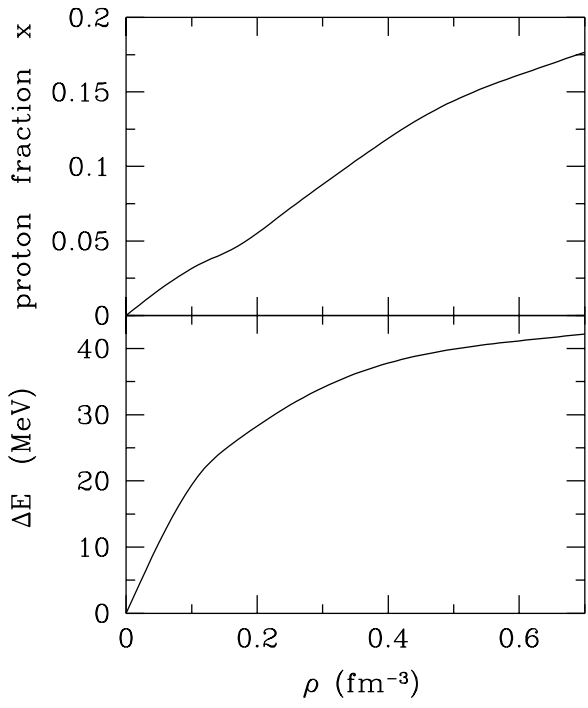,height=10cm}
            \epsfig{file=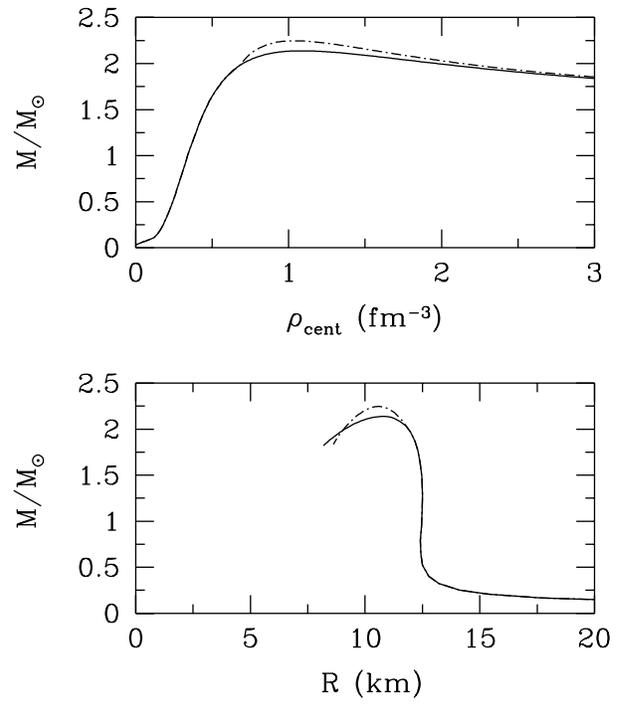,height=10cm}}
\caption{Left panel: proton fraction and $\Delta E$.
Right panel: nucleon star properties.
}
\label{fignstar}
\end{figure}

In this section we apply the nuclear matter equation of state
obtained in the DBHF calculation to the study of neutron star
properties.   In a real neutron star, there exists a large fraction of
protons with electrons and muons maintaining the charge neutrality.
We call this star a nucleon star instead of neutron star \cite{Geb97}.
We assume that the star is in chemical equilibrium at zero 
temperature after the neutrinos leave the system. 
The equilibrium condition requires that the  chemical
potentials should satisfy
\be
\hat\mu =\mu_n-\mu_p=\mu_e=\mu_\mu,
\label{eqEQ}\ee
where muons start to contribute when $\mu_e > m_\mu$.
The nucleon chemical potential difference
$\hat\mu$ can be calculated once we have the coefficients of symmetry 
energy from DBHF \cite{Pralec},
\be
\hat\mu =4 (1-2x) \left[ S_2(\rho)+2 S_4 (\rho) (1-2x)^2\right],
\label{eqmuhat}\ee
where $x$ being the proton fraction. 
{}From the charge neutrality $\rho_e+\rho_\mu=\rho_p$ and Eq. (\ref{eqEQ}),
we have 
\be
\rho x =\frac 1{3\pi^2} \mu_e^3 +
\theta(\mu_e-m_\mu) \frac{1}{3\pi^2} \left(\mu_e^2-m_\mu^2 \right)^{3/2}.
\label{eqbetaeq}\ee
By solving Eqs. (\ref{eqmuhat}) and (\ref{eqbetaeq})
for a given density, we can get the proton
fraction in nucleon star matter.
In Fig. \ref{fignstar}, the proton fraction and the energy difference
(Eq. (\ref{eqde}))
are plotted in the left panel. Muons start to contribute almost at the
nuclear saturation density, $\rho_0=0.16$ fm$^{-3}$, which leads to
a slight bump in the proton fraction.

Given the symmetry energy and EOS of symmetric nuclear matter, we can
get the energy density and pressure and solve the 
Tolman-Oppenheimer-Volkov (TOV)
equations for a spherically symmetric nucleon star, 
\be
\frac{dM}{dr} &=& 4\pi r^2 \epsilon \nonumber\\
\frac{dP}{dr} &=& 
 - \frac{G \left[\epsilon+ P\right]\left[1+ 4\pi r^3 P\right]}{r(r-2GM)}.
\ee
In the above equations, the pressure 
$P(\epsilon)$ determines the stiffness of the TOV equation,
On the right panel of Fig. \ref{fignstar}, the nucleon star properties
are summarized, where the dash-dotted curves give the results 
without inclusion of muons. The maximum mass nucleon star corresponds to 
$M_{{\rm max}}=2.1\; M_\odot$, $R=10.8$ km, and the central density
$\rho_c = 1.08$ ${\rm fm}^{-3}$. This mass, we believe, is well 
determined in our conventional scenario, beginning from the 
microscopic Bonn two-body interactions, and using the many-body 
theory as described above. The EOS can be substantially softened, 
however, by introduction of kaon condensation \cite{TPL,BLRT} and this
brings the maximum mass down to $M_{{\rm max}}\sim 1.5\; M_\odot$ \cite{BB}.
Introduction of $K^-$-mesons implies an $\sim 50 \%$ nucleus 
components of protons at the higher densities, so that "nucleon" star,
rather than neutron star, then becomes the most appropriate name.
A recent compilation by Steve Thorsett quoted by Brown \cite{Geb97}
shows that well-measured neutron star masses are all less than $1.5\; M_\odot$.

On the other hand, within the nuclear many-body framework, 
Jiang {\it et al.} \cite{jiang93}
have shown that the inclusion of ring diagrams in the DBHF 
calculations softens the equation of state of symmetric nuclear matter.
It would be of interest to extend this to
asymmetric nuclear matter, and to see to what entent this
improvement in the nuclear many-body approach can improve the
agreement with the astrophysics observations.

\section{Summary}

In summary, we studied the properties of asymmetric nuclear matter
in the formalism of the Dirac-Brueckner approach with 
the Bonn one-boson-exchange
nucleon-nucleon interaction. The symmetry energy coefficient
at the saturation density obtained in this work is about
30 MeV. This is in good agreement with the empirical value of
about 34$\pm 4$ MeV, and in agreement with other approaches
such as the BHF \cite{bom91} and the variational \cite{wir88}
calculations. The symmetry energy in our study was found to 
increase almost linearly with the density and agrees with the
linear parameterization of Prakash, Ainsworth and Lattimer 
\cite{pra88}.  At higher densities, the symmetry energy in
our calculation is considerably larger than those in
the BHF and variational calculations. 
The difference can be understood as coming from the
both the relativistic effects in the 'kinetic energy' contribution,
and a strong $\rho$-meson coupling in the Bonn potential
that increases the 'potential energy' contribution,
to the symmetry energy.

We have also applied the resulting equation
of state of neutron-rich matter to calculate the
maximum mass of nucleon star, and we find it to be about 2.1$M_\odot$. 
The corresponding radius and central density are $R=10.8$ km and
$\rho_c=1.08$ fm$^{-3}$, respectively. 
This maximum mass is substantially greater than that measured in any
neutron stars. It would be brought down considerably if kaon
condensation were included in the EOS \cite{TPL,BB}.
We believe our maximum mass of 2.1 $M_\odot$, without inclusion of kaon
condensation, to be the most direct determination of this important
parameter, since we begin from a microscopic two-body interaction and use
the best presently available technology to calculate the mass.

\vskip 1cm
\section*{Acknowledgement}

We would like to thank M. Prakash, M. Rho and C.M. Ko for 
helpful discussions. 
We also appreciate R. Machleidt for sending us the code of the 
relativistic Bonn potential.
This work was supported by the U.S. Department of Energy under 
grant no. DE-FG02-88ER40388.
The work of CHL was supported in part by Korea Science and 
Engineering Foundation

\newpage

\newpage

\newpage

\newpage

\end{document}